\documentclass[a4paper,twoside]{article}

\usepackage{epsfig}
\usepackage{hyperref}
\usepackage{subcaption}
\usepackage{calc}
\usepackage{amssymb}
\usepackage{amstext}
\usepackage{amsmath}
\usepackage{amsthm}
\usepackage{multicol}
\usepackage{pslatex}
\usepackage{apalike}
\usepackage{physics}
\usepackage{algorithm2e}
\usepackage[bottom]{footmisc}
\usepackage{SCITEPRESS}     % Please add other packages that you may need BEFORE the SCITEPRESS.sty package.

\begin{document}

\title{Quantum-Efficient Kernel Target Alignment}

\author{\authorname{Rodrigo Coelho\sup{1}, Georg Kruse\sup{1}\sup{2} and Andreas Rosskopf\sup{1}}
\affiliation{\sup{1}Fraunhofer IISB, Erlangen, Germany}
\affiliation{\sup{2}Technical University Munich}
\email{\{rodrigo.coelho, georg.kruse, andreas.rosskopf\}@iisb.fraunhofer.de}
}


\abstract{In recent years, quantum computers have emerged as promising candidates for implementing kernels. Quantum Embedding Kernels embed data points into quantum states and calculate their inner product in a high-dimensional Hilbert Space by computing the overlap between the resulting quantum states. Variational Quantum Circuits (VQCs) are typically used for this end, with Kernel Target Alignment (KTA) as cost function. The optimized kernels can then be deployed in Support Vector Machines (SVMs) for classification tasks. However, both classical and quantum SVMs scale poorly with increasing dataset sizes. This issue is exacerbated in quantum kernel methods, as each inner product requires a quantum circuit execution. In this paper, we investigate KTA-trained quantum embedding kernels and employ a low-rank matrix approximation, the Nyström method, to reduce the quantum circuit executions needed to construct the Kernel Matrix. We empirically evaluate the performance of our approach across various datasets, focusing on the accuracy of the resulting SVM and the reduction in quantum circuit executions. Additionally, we examine and compare the robustness of our model under different noise types, particularly coherent and depolarizing noise.}

\onecolumn \maketitle \normalsize \setcounter{footnote}{0} \vfill

\section{\uppercase{Introduction}}
\label{sec:introduction}

Quantum computers may potentially solve certain problems faster than classical computers. However, in the \emph{Noisy Intermediate Scale Quantum (NISQ)} era, we are limited in the algorithms that quantum computers can implement \cite{preskill2018quantum}. Consequently, algorithms that theoretically offer advantages over the best-known classical methods, such as Shor's \cite{shor1999polynomial} and Grover's \cite{grover1996fast} algorithms, cannot yet be implemented to tackle problems of industrial significance. In the NISQ era, quantum computing has focused on Variational Quantum Circuits (VQCs). These circuits rely on free parameters that are iteratively updated by a classical optimizer. VQCs are suitable for NISQ devices due to their low requirements in both number of qubits and circuit depth, which mitigates noise effects \cite{cerezo2021variational}. Typically used as function approximators, VQCs are the quantum analog of Neural Networks (NNs), as both are black-box models that depend on parameters iteratively adjusted to minimize a cost function \cite{abbas2021power}. Given their potential, VQCs have been extensively applied in machine learning and form a significant component of Quantum Machine Learning (QML). Notable examples include the Quantum Approximate Optimization Algorithm (QAOA) \cite{farhi2014quantum} for solving combinatorial optimization problems, the Variational Quantum Eigensolver (VQE) \cite{kandala2017hardware} for finding ground states of Hamiltonians, and their application in both supervised \cite{schuld2018supervised} and reinforcement \cite{skolik2022quantum} learning.

On the classical side, kernel methods are one of the cornerstones of machine learning, known for their effectiveness in handling non-linear data by using implicit feature spaces. These methods map input data into a higher-dimensional space where linear separation is possible, facilitated by the kernel trick, which computes inner products in this space without explicitly performing the transformation. Support Vector Machines (SVMs) are a prime example, leveraging kernel methods to find optimal decision boundaries for classification tasks \cite{hearst1998support}. Even though we'll focus on SVMs throughout this work, kernel methods can be applied to many tasks beyond classification, ranging from regression \cite{drucker1996support} to clustering \cite{dhillon2004kernel}.

In recent years, interest in exploring how quantum computing can enhance kernel methods has increased. Quantum computers, can naturally process high-dimensional spaces, providing a promising platform for implementing kernels. For this reason, they have been extensively explored recently \cite{wang2021towards}\cite{jager2023universal}.We are interested in Quantum Embedding Kernels (QKEs), which use quantum circuits to embed data points into a high-dimensional Hilbert space. The overlap between these quantum states is then used to compute the inner product between data points in this feature space. Typically, these kernels are parameterized, making them a form of VQCs. The parameters of these circuits are optimized based on \emph{Kernel Target Alignment (KTA)}, which serves as a metric to align the kernel with the target task (which we will consider to be binary classification) \cite{hubregtsen2022training}. Once the kernel is optimized, it is fed into an SVM to determine the optimal decision boundary for the classification task at hand. This integration leverages the strengths of both quantum and classical computing, aiming to enhance classification performance due to the expressive power of quantum embeddings combined with the robust framework of SVMs.

However, the method scales poorly. For instance, using the KTA as cost function requires calculating the kernel matrix at every single training step, a process that scales quadratically $O(N^2)$ with training dataset size $N$. To alleviate this, both the original paper \cite{hubregtsen2022training} and a following one \cite{sahin2024efficient} propose using only a subsample of size $D\ll N$ of the training dataset to compute the KTA at each epoch, effectively turning the computation, which becomes $O(D^2)$, independent of $N$. Moreover, another paper \cite{tscharke2024quack} proposes using a clustering algorithm to find centroids of the classes and then computing the kernel matrix with respect to these centroids at each training step, bringing the complexity of the computation to $O(N)$. So, these methods are able to decrease the complexity of training from scaling quadratically with $N$ to either scaling only linearly or even being completely independent of $N$. Nonetheless, the end goal is for the optimized kernel matrix, which contains the pairwise inner-products between all training data points, to be fed into the SVM. Thus, at least for this final computation, these methods still require $O(N^2)$ computations. This is especially important if we consider that each pairwise inner-product requires a quantum circuit execution.

In this paper, we adapt a low-rank matrix approximation known as the \emph{Nyström Approximation}, which is commonly used in classical kernel methods, to address this scalability issue. This approach reduces the complexity of computing the kernel matrix from $O(N^2)$ to O($NM^2$), where $M\ll N$ is an hyperparameter that determines the quality of the approximation. The Nyström Approximation is applied for computing the optimized kernel matrix that is fed into the SVM, resulting in a classification pipeline that scales linearly with the training dataset size $N$ in all steps. Consequently, our method facilitates the efficient application of quantum kernel methods to industrially-relevant problems.

The contributions of this paper are organized as follows:
\begin{itemize}
    \item Adapted the Nyström Approximation to Quantum Embedding Kernels
    \item Empirically verified (on several synthetic datasets) that, in a noiseless setting, the Nyström Approximation method allows for quantum kernels with reduced quantum circuit executions at a small cost in the accuracy of the resulting SVM.
    \item Empirically tested the performance of the developed method under both coherent and incoherent noise.
\end{itemize}

\section{\uppercase{Kernel Methods and Support Vector Machines}}

Kernel methods can be used for different tasks, from dimensionality reduction \cite{scholkopf1997kernel} to regression \cite{drucker1996support}. We will focus on binary classification using SVMs, which are linear classifiers \cite{hearst1998support}. Nonetheless, they can be used in non-linear classification problems due to the kernel trick, which implicitly maps the data into high-dimensional feature spaces where linear classification is possible. In this section we will go through this pipeline for classification, starting with kernel methods and ending with SVMs.

\subsection{\uppercase{Kernel Methods}}

Consider a dataset $X=\{(x_i,y_i)\}^n_{i=1}$, where $x_i\in\mathbb{R}^d$, with $d$ being the number of features of $x_i$. A kernel method involves defining a feature map $\phi:\mathbb{R}^d\rightarrow \mathcal{H}$, where $\mathcal{H}$ is a high-dimensional Hilbert space. The kernel function can then be defined as:

\begin{equation}\label{kernel_function}
    k(x_i, x_j) = \langle \phi(x_i), \phi(x_j) \rangle_{\mathcal{H}}
\end{equation}

Put in words, the kernel function computes the inner product between the inputs $x_i$ and $x_j$ in some high-dimensional feature space $\mathcal{H}$. Given the kernel function $k$, the kernel matrix $K$ is a symmetric matrix that contains the pairwise evaluations of $k$ over all the points in the training dataset $X$:

\begin{equation}
    K_{ij} = K(x_i,x_j) = \langle \phi(x_i), \phi(x_j) \rangle_{\mathcal{H}}
\end{equation}

Typically, explicitly calculating the coordinates of the input points in the high-dimensional feature space, a task that may be computationally expensive, is not needed. Instead, the \emph{kernel trick} allows one to efficiently compute these inner products without needing to explicitly calculate the feature map $\phi$.

For example, consider the Radial Basis Function (RBF) kernel, defined as:

\begin{equation}
    k_{RBF}(x, y) = \exp(-\gamma \|x - y\|^2)
\end{equation}

where $\gamma$ is an hyperparameter that the user defines. This kernel computes the inner product between the data points $x$ and $y$ in an infinite-dimensional feature space given by the feature map $\varphi_{RBF}$. However, due to the kernel trick, one only needs to compute the kernel $k_{RBF}$ and not the feature map $\varphi_{RBF}$, saving computational resources.

\subsection{\uppercase{Support Vector Machines}}

An SVM aims to find the optimal hyperplane that separates data points of two different classes with the maximum margin. This margin is the distance between the hyperplane and the nearest data points from either class, known as support vectors \cite{hearst1998support}. SVMs are inherently linear classifiers; they work effectively when the dataset is linearly separable by a hyperplane.

However, the kernel trick extends SVMs to non-linear datasets. By using a kernel function, the data is implicitly mapped into a high-dimensional feature space where linear separation is feasible. Thus, with the aid of the kernel trick, SVMs can classify non-linear datasets.

Given a kernel matrix $K$ containing the pairwise inner-products between all training points $x\in X$, the optimization problem the SVM solves is formulated as:

\begin{equation}\label{svm_optim}
    \min_{\alpha} \frac{1}{2} \sum_{i=1}^{n} \sum_{j=1}^{n} \alpha_i \alpha_j y_i y_j K(x_i, x_j) - \sum_{i=1}^{n} \alpha_i
\end{equation}

where $\alpha_i$ are the Lagrange multipliers. An SVM solves a quadratic problem and thus is guaranteed to converge to the global minimum. 

The decision function is given by:

\begin{equation}
    f(x) = sign\left(\sum_{i=1}^n \alpha_i y_i K(x_i,x) + b\right)
\end{equation}

Here, $b$ is the bias term, often determined using the support vectors. However, as shown in Equation \ref{svm_optim}, training an SVM on a dataset of size $N$ requires computing $N^2$ pairwise inner products to construct the kernel matrix $K$. This implies that the runtime scales at least quadratically with the training dataset size $N$. Moreover, even after obtaining the kernel matrix $K$, the SVM must solve a quadratic optimization problem. Depending on the data structure and the algorithm used, this process may scale with $N^2$ or even $N^3$. Consequently, SVMs are typically applied to small-scale problems with moderate dataset sizes.

\begin{figure}[t]
    \centering
    \includegraphics[width=\linewidth]{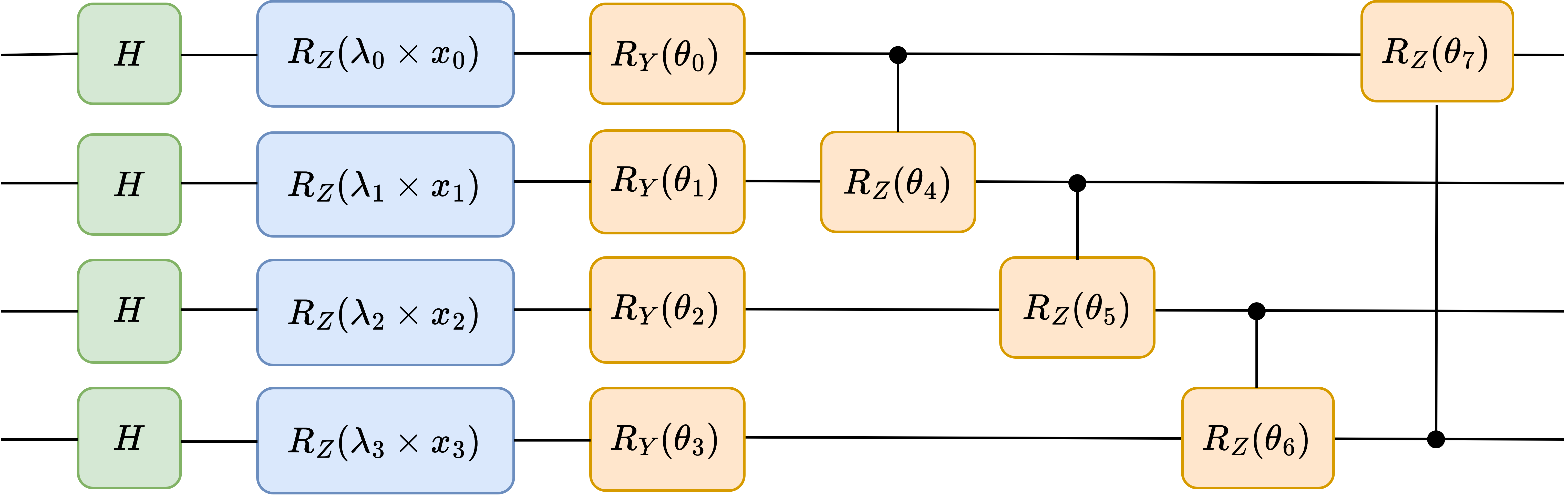}
    \caption{One layer of the quantum ansatz used throughout this work. As an example, this layer in particular contains $4$ qubits. The ansatz contains input scaling parameters $\lambda$ and variational parameters $\theta$. The data encoding gates (blue-colored) are repeated in every single layer - data re-uploading.}
    \label{fig:quantum_ansatz}
\end{figure}

\begin{figure*}[t]
    \centering
    \begin{subfigure}[b]{0.48\textwidth}
        \centering
        \includegraphics[width=\textwidth]{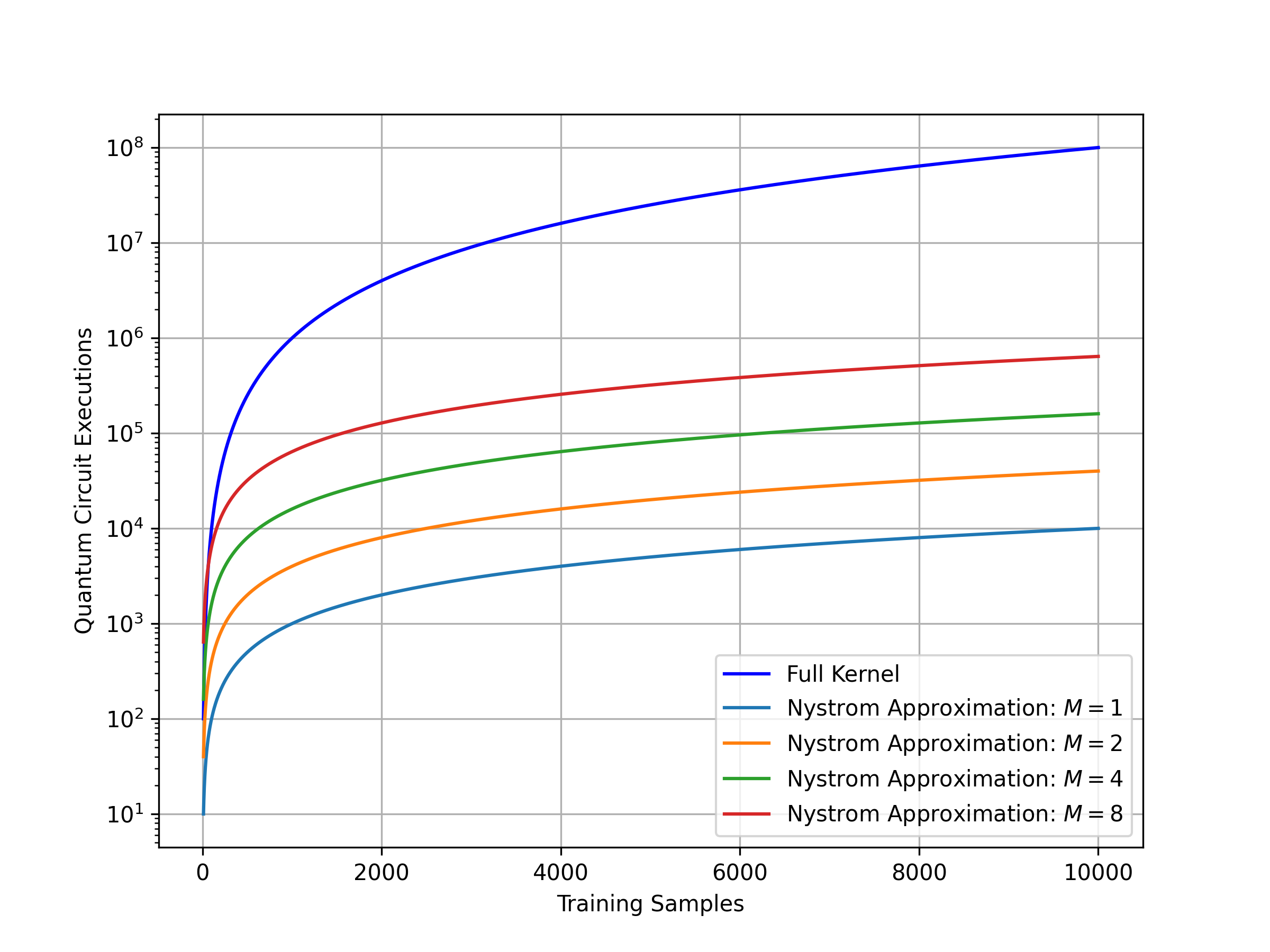}
        \caption{}
        \label{fig:training_executions}
    \end{subfigure}
    \hspace{0.02\textwidth}
    \begin{subfigure}[b]{0.48\textwidth}
        \centering
        \includegraphics[width=\textwidth]{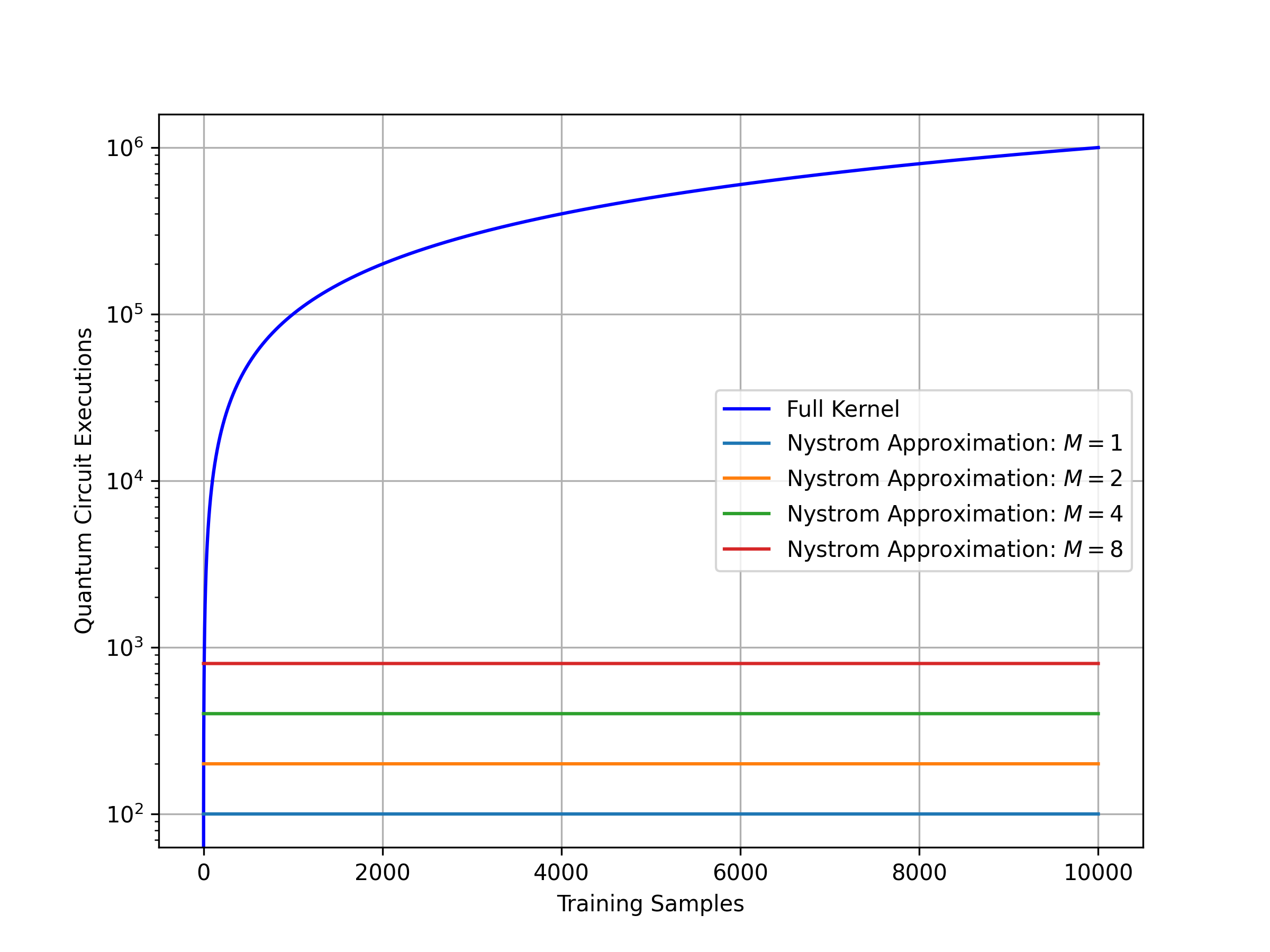}
        \caption{}
        \label{fig:testing_executions}
    \end{subfigure}
    \caption{Quantum circuit executions required for computing the training kernel matrix (Fig.\ref{fig:training_executions}) and testing kernel matrix (Fig.\ref{fig:testing_executions}) as a function of the training dataset size $N$ for the standard approach and the Nyström approximation with different $M$s.}
    \label{fig:circuit_executions}
\end{figure*}

\section{\uppercase{Quantum Embedding Kernels}}

The first step to solve a classical problem using a quantum computer is to encode the classical data into a quantum state that can be processed by quantum operations. Given a data point $x$, one can define a quantum circuit $U(x)$ to generate said quantum state:
\begin{equation}
    \ket{\varphi(x)} = U(x)\ket{0}
\end{equation}

This corresponds to embedding the data in a high-dimensional Hilbert space. Referring back to Equation \ref{kernel_function}, a kernel function is defined as the inner product between two data points in a high-dimensional Hilbert space. In the context of quantum computing, this translates to calculating the fidelity between the quantum states generated by encoding the two data points. The fidelity, which measures the similarity between these quantum states, is given by:

\begin{equation}
    k(x_i,x_j) = \left| \langle \varphi(x_i) | \varphi(x_j) \rangle \right|^2
\end{equation}

There are several ways to calculate the fidelity between two quantum states. In particular, assuming that $\varphi(x)$ and $\varphi(y)$ are pure quantum states (that is, $Trace(\rho^2)=1$, where $\rho=\bra{\varphi}\ket{\varphi}$), then the fidelity can be calculated using the adjoint of the quantum circuit \cite{hubregtsen2022training}:

\begin{equation}\label{quantum_kernel}
    K(x,y) = \left| \langle \varphi(x) | \varphi(y) \rangle \right|^2 = \left| \langle 0 | U^\dagger(y) U(x) | 0 \rangle \right|^2
\end{equation}

This is equivalent to the probability of observing $\ket{0}$ when measuring $U^T(y)U(x)\ket{0}$ in the computational basis.

We have established a definition for quantum kernels. Nevertheless, the choice of which quantum kernel to use remains unclear. In Noisy-Intermediate Scale Quantum (NISQ) devices, a popular approach is to make the circuit trainable, define a cost function, and optimize the parameters using a classical optimizer. These methods are known as Variational Quantum Algorithms. In our specific case, a quantum kernel can be made trainable by making the quantum circuit $U(x)$ also depend on these classical parameters, becoming $U(x,\theta)$. Thus, the kernel function from Eq. \ref{quantum_kernel} becomes:

\begin{equation}
    K_\theta(x,y) = \left| \langle 0 | U^\dagger(y,\theta) U(x,\theta) | 0 \rangle \right|^2
\end{equation}

Then, the question becomes what cost function to use. Following paper \cite{hubregtsen2022training}, we use \emph{Kernel-Target Alignment (KTA)} as our cost function, which predicts the alignment between the quantum kernel $K_\theta$ and the labels of the training data. We start by defining an ideal kernel using the labels of the training data:

\begin{equation}
    k_y(x_i,x_j) = y_iy_j
\end{equation}

Then, assuming $y\in\{-1,1\}$, $k_y$ will be $1$ if both labels belong to the same class and $-1$ otherwise. This ideal kernel matrix can be defined as:

\begin{equation}
    K_y = y^Ty
\end{equation}

Then, the KTA can be defined as:

\begin{equation}
    KTA=\frac{y^TKy}{\sqrt{Tr(K^2)}N}
\end{equation}

In this context, $N$ represents the total number of training samples in the dataset. Since we want to minimize a cost function and maximize KTA, we will include a negative sign in this equation during training.

\begin{figure*}[ht]
    \centering
    \includegraphics[width=\textwidth]{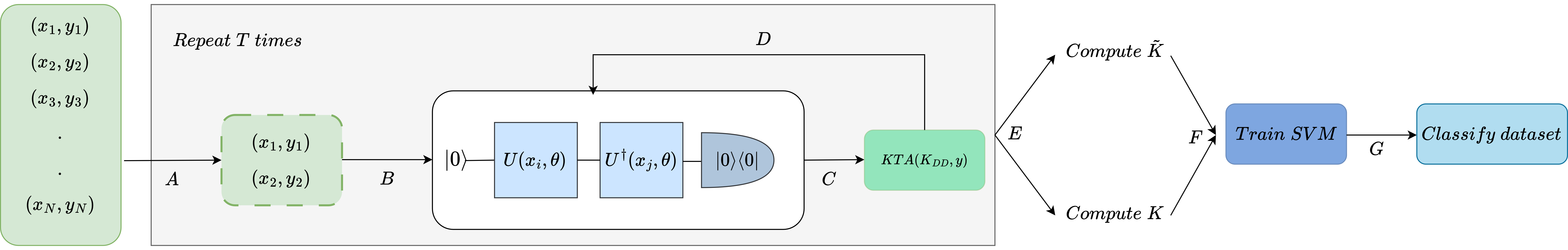}
    \caption{Given a training dataset of size $N$, for $T$ iterations, do: A) sample a random mini-batch of data points of size $D$. B) Calculate the overlap between the embedded quantum states. C) Once all the pairs have been processed, fill the kernel matrix $K_{DD}$ and compute the cost function (KTA). D) Update the parameters $\theta$ of the quantum circuit. Finally, after the $T$ training iterations, E) Compute the approximated training kernel matrix $\Tilde{K}$ using either the Nyström approximation method or the full training kernel matrix $K$ using the standard approach. F) Use the training kernel matrix to fit an SVM to the dataset and G) Classify the training dataset. }
    \label{fig:kernel_pipeline}
\end{figure*}

\section{\uppercase{Nyström Approximation}}

To reduce the complexity of computing the kernel matrix, one can use the Nyström approximation \cite{drineas2005nystrom}. It works as follows. The first step is to select a subset of $M\ll N$ data points from the original training dataset $N$, referred to as landmarks. In this work, we will always select the landmarks randomly, but they can also be selected according to more sophisticated strategies, such as K-means clustering. Then, compute the matrix $K_{MM}$ of shape $(M,M)$ using the selected subset where, given landmarks $\{m_1,m_2,...,m_M\}$, $K_{MM}(i,j)=\langle \phi(m_i), \phi(m_j) \rangle_{\mathcal{H}}$. Then, compute the cross-kernel matrix $K_{NM}$ of shape $(N,M)$ between the $M$ landmarks and the $N$ training data points. Finally, the Nyström approximation of the kernel matrix $K$ is given by:

\begin{equation}
    \Tilde{K}\approx K_{NM}K^{-1}_{MM}K^{T}_{NM}
\end{equation}

If we disregard the computational cost of inverting $K_{MM}$ (as this step is performed classically), the Nyström approximation reduces the computational cost of calculating the kernel matrix from $O(N^2)$ to $O(NM^2)$. Consequently, the number of quantum circuit executions required to compute the matrix now scales linearly with $N$, significantly enhancing efficiency in the quantum context.

This method can also be used to reduce the computational cost of generating the kernel matrix to classify a test dataset. The end goal in classification tasks is to test the model on unseen data points. With an SVM, this requires generating a testing kernel matrix of shape $(P,N)$, where $P$ is the size of the testing dataset, leading to a computational complexity of $O(PN)$. However, using the Nyström approximation, one can reduce this computational cost to $O(PM)$, making it independent of the training dataset size $N$. This is accomplished by computing:

\begin{equation}
\Tilde{K}_{\text{test}} \approx K_{PM}^T K_{MM}^{-1} K_{NM}
\end{equation}

Here, $K_{PM}$ is the cross-kernel matrix between the $M$ landmarks and the $P$ test data points, and $K_{MN}$ is the previously computed cross-kernel matrix between the $M$ landmarks and the $N$ training data points. For a comparison in terms of the number of quantum circuit executions required to construct the training and testing kernel matrices using either the Nyström approximation or the standard approach, see Fig.\ref{fig:circuit_executions}

\section{\uppercase{Method}}

In this work, we test the Nyström approximation method for generating the kernel matrix that is fed into the SVM in the context of quantum embedding kernels. We start with a training dataset with $N$ data points and an ansatz for the VQC, see Fig. \ref{fig:quantum_ansatz}, randomly initializing the free parameters.  The quantum circuit is then trained over $T$ training iterations using the KTA as the cost function. 

However, calculating the kernel matrix $K$ for every training iteration is computationally expensive. To mitigate this, we adopt a strategy similar to that used in \cite{hubregtsen2022training,sahin2024efficient}. At each iteration, we randomly sample a mini-batch of $D$ training points, construct their kernel matrix $\Tilde{K}_{DD}$, calculate the KTA for this mini-kernel, and update the parameters using a classical optimizer. Consequently, the number of quantum circuit executions per training iteration becomes independent of the size of the training dataset $N$ and instead scales quadratically with $D$. Since, as we will see, $D$ can be chosen such that $D\ll N$, this method significantly reduces the number of quantum circuit executions required per training iteration.

After completing the $T$ training iterations and optimizing the quantum kernel for the task at hand, we compute the kernel matrix for the entire training dataset $N$ using the Nyström approximation method. This is our main contribution in this work. Using this approximation allows us to build the kernel matrix that is fed into the SVM at the end of training using only $O(NM^2)$ (where $M$ is the number of landmarks), quantum circuit executions, instead of the $O(N^2)$ executions that both \cite{hubregtsen2022training} and \cite{sahin2024efficient} require. To our knowledge, this is the first pipeline in which the number of quantum circuit executions scales linearly with the training dataset size $N$ in all steps. Specifically, the training process has a complexity of $O(D^2)$, and the computation of the final kernel matrix of $O(NM^2)$ (note that this complexity takes into account only the number of quantum circuit executions). The entire pipeline can be seen in Fig.\ref{fig:kernel_pipeline}. 

Although not explicitly shown in the figure, the Nyström approximation method saves quantum circuit executions in comparison with the standard approach during inference as well, when classifying a testing dataset. The standard approach has a (quantum) complexity of $O(NP)$, with $P$ being the size of the testing dataset, while the Nyström approximation has a complexity of $O(PM)$.

\begin{figure*}[ht]
    \centering
    \begin{subfigure}[b]{0.48\textwidth}
        \centering
        \includegraphics[width=\textwidth]{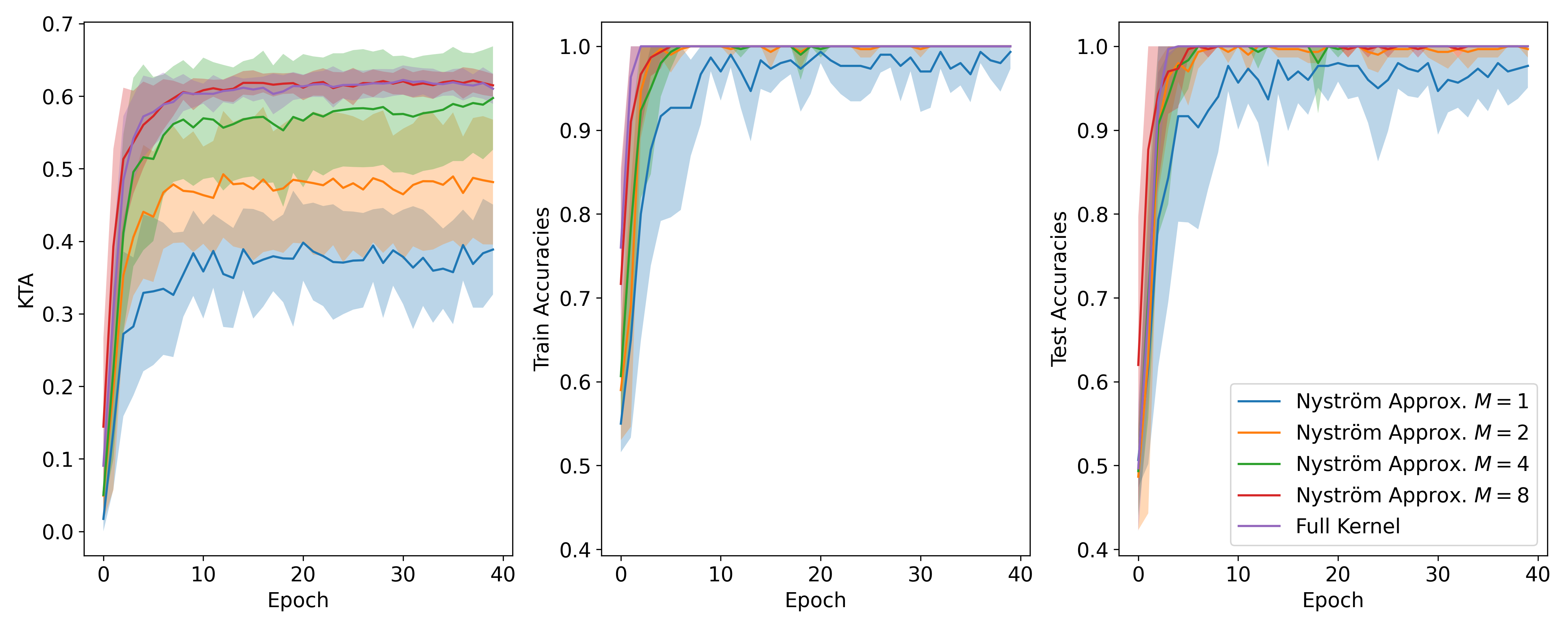}
        \caption{}
        \label{fig:subfig1}
    \end{subfigure}
    \hfill
    \begin{subfigure}[b]{0.48\textwidth}
        \centering
        \includegraphics[width=\textwidth]{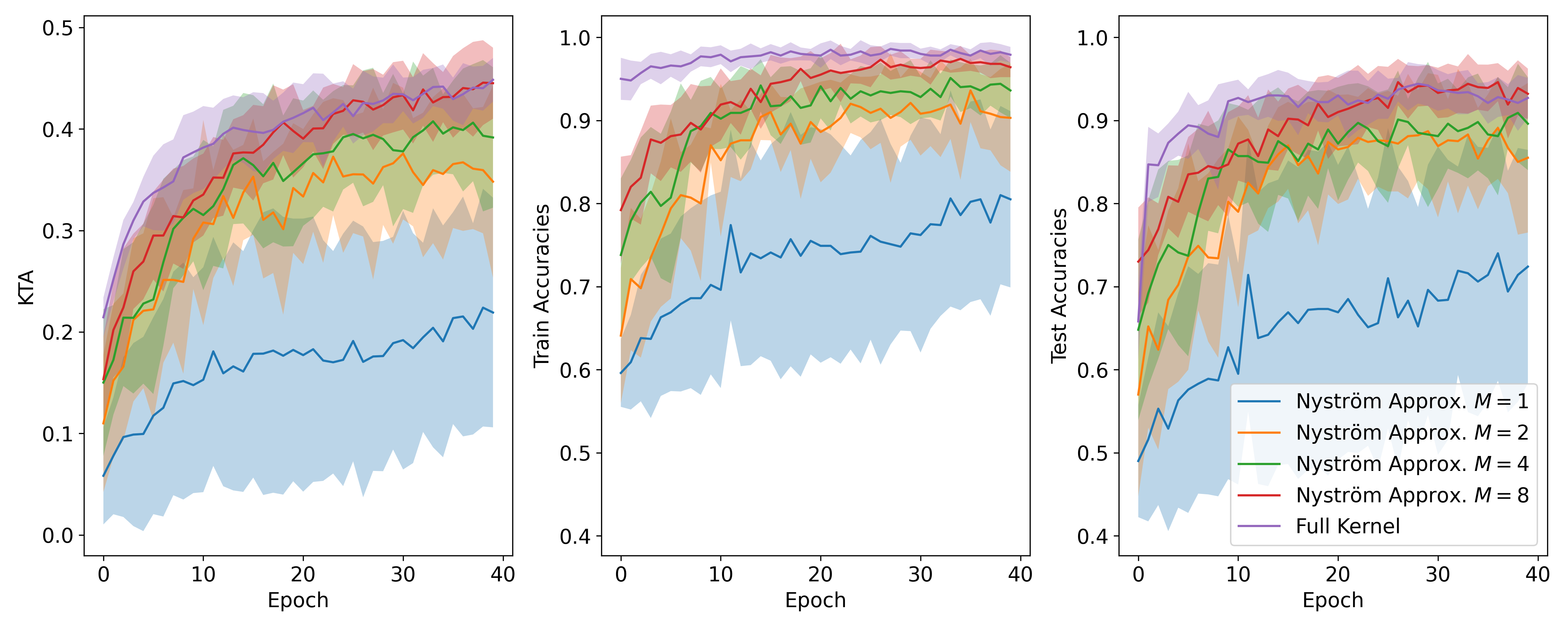}
        \caption{}
        \label{fig:subfig2}
    \end{subfigure}
    \\[1em] % Adjust the space here (1.5em is an example)
    \begin{subfigure}[b]{0.48\textwidth}
        \centering
        \includegraphics[width=\textwidth]{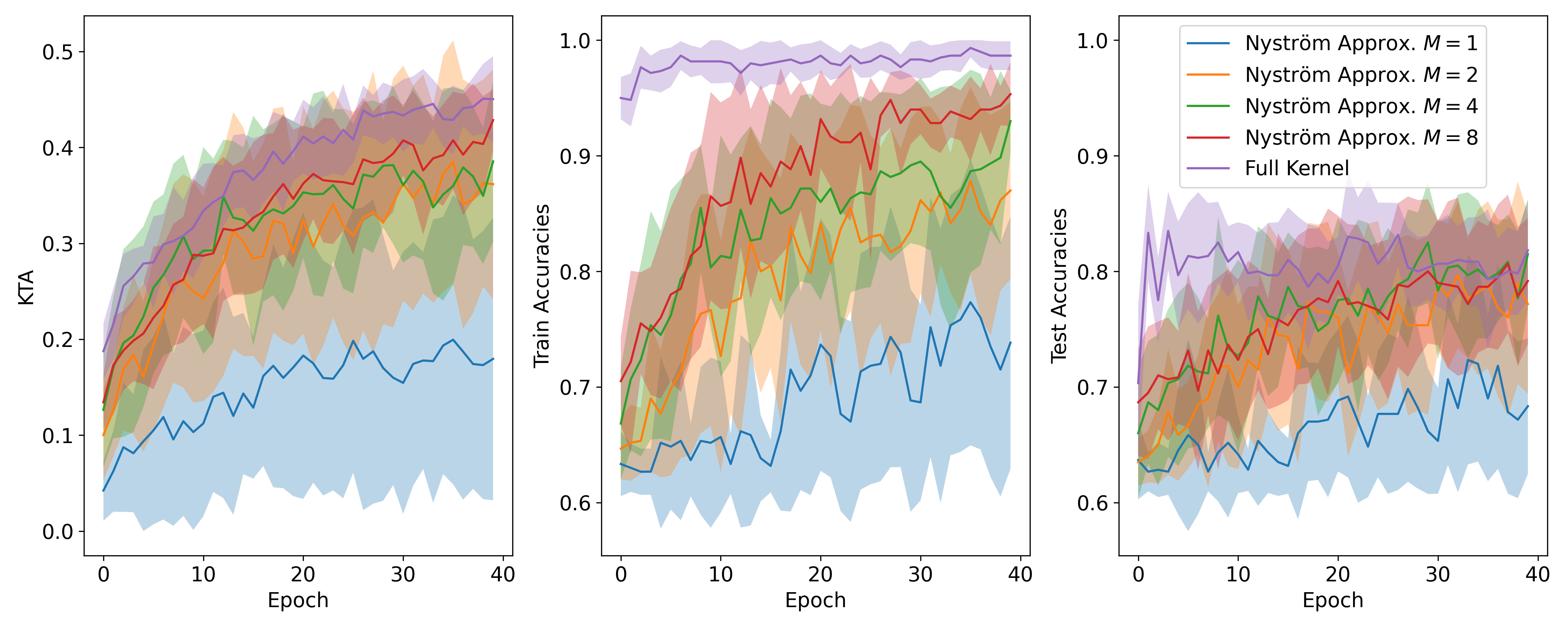}
        \caption{}
        \label{fig:subfig3}
    \end{subfigure}
    \hfill
    \begin{subfigure}[b]{0.48\textwidth}
        \centering
        \includegraphics[width=\textwidth]{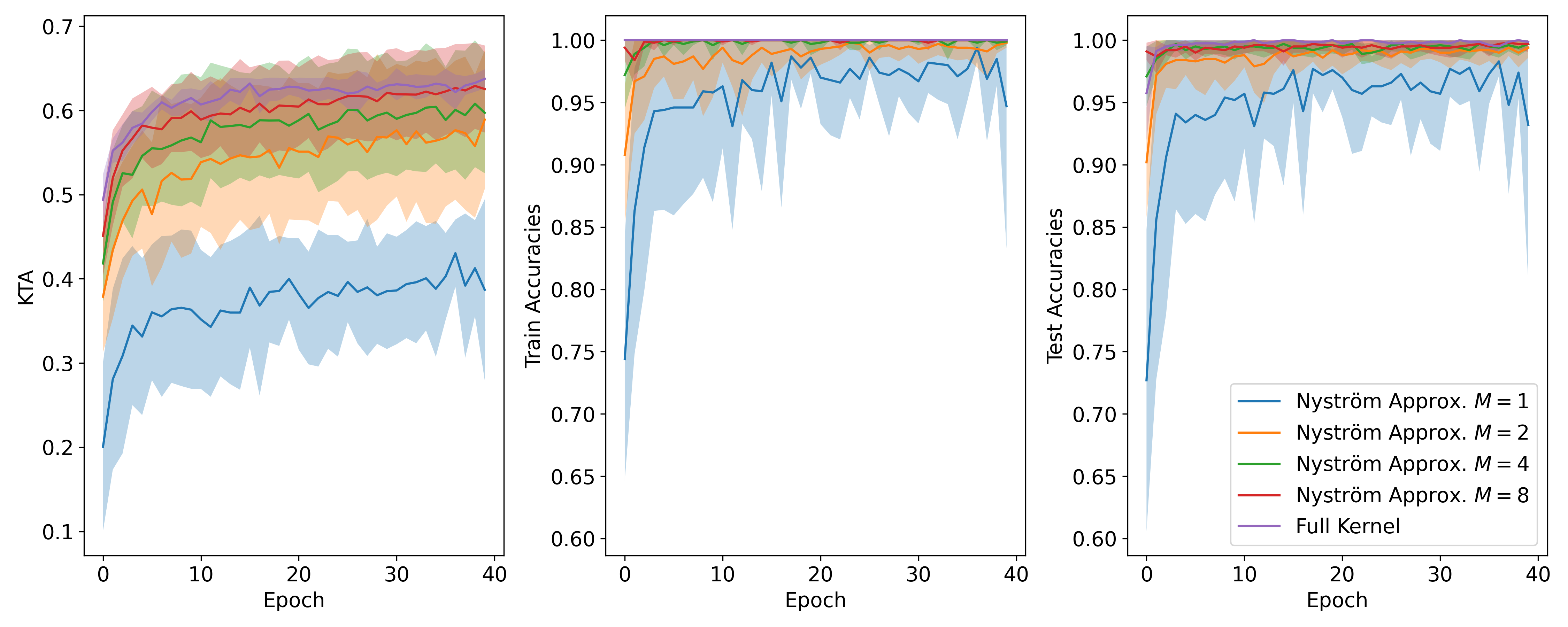}
        \caption{}
        \label{fig:subfig4}
    \end{subfigure}
    \caption{Performance of the Standard approach VS the Nyström approximation with different $Ms$ throughout training measured through KTA, training and testing accuracy in the checkers (Fig. \ref{fig:subfig1}), corners (Fig. \ref{fig:subfig2}), donuts (Fig. \ref{fig:subfig3}) and spirals (Fig. \ref{fig:subfig4}) datasets.}
    \label{fig:noiseless_results}
\end{figure*}

Lastly, we wish to explain in higher depth the differences between the method from \cite{sahin2024efficient} and the Nyström approximation method we propose here. In short, they are different methods with different applications. The method from \cite{sahin2024efficient} allows us to use KTA as the cost function without computing a full kernel matrix at each training step. Instead, it computes a mini-kernel matrix $K_{DD}$, reducing quantum circuit executions from $O(N^2)$ to $O(D^2)$, where $D \ll N$. This is essentially a type of stochastic gradient descent, where KTA is approximated by using a subset of data points.

After training, the SVM requires the full optimized kernel matrix to classify the training dataset. Therefore, we cannot use the previous method here, as the SVM needs the complete matrix ($O(N^2)$ quantum circuit executions) rather than a mini-version based on a subsample. The Nyström Approximation method addresses this need by computing an approximated full-kernel matrix $\Tilde{K}$ while reducing the number of quantum circuit executions to $O(NM^2)$. Thus, using both methods —\cite{sahin2024efficient} during training and the Nyström Approximation for generating the kernel matrices once training is done —results in a pipeline that scales linearly with the training dataset size in all steps.

We can also apply the Nyström Approximation during the $T$ training iterations instead of the method from \cite{sahin2024efficient} by using the approximated kernel matrix $\Tilde{K}$ to compute KTA. However, this does not reduce quantum circuit executions and, based on limited experiments, did not yield significantly better results. Therefore, in this work, we use the method from \cite{sahin2024efficient} during training.

\begin{figure*}[ht]
    \centering
    \begin{subfigure}[b]{\textwidth}
        \centering
        \includegraphics[width=0.8\textwidth]{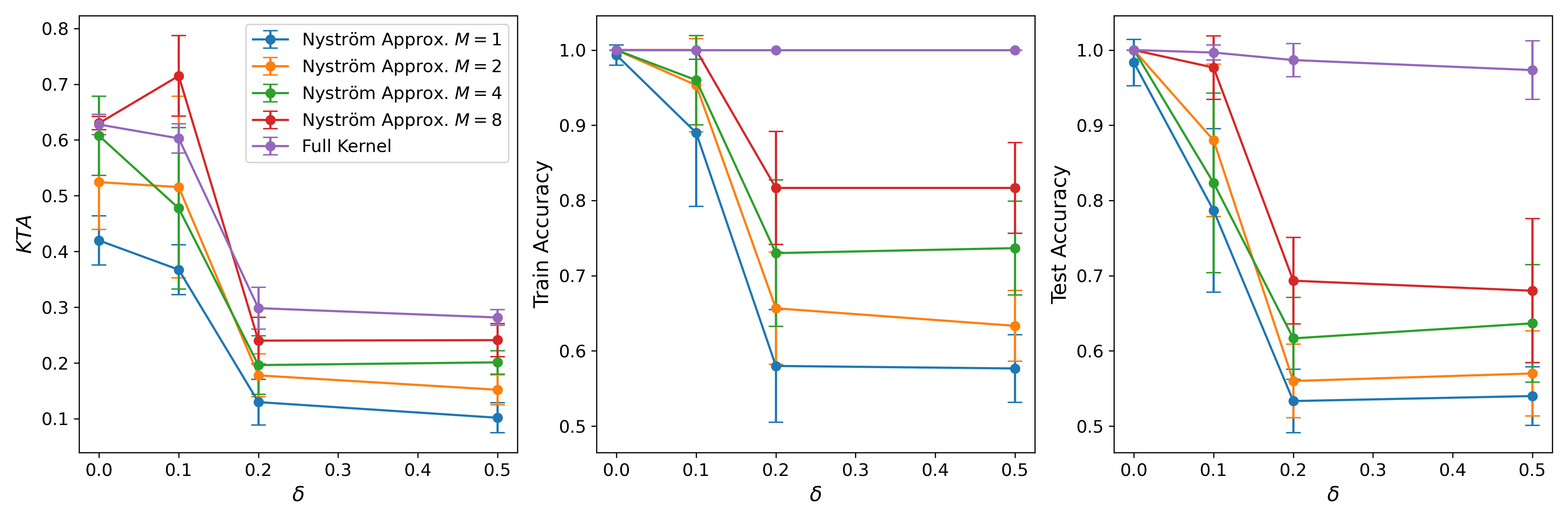}
        \caption{}
        \label{fig:cohr1}
    \end{subfigure}
    \\[1em]
    \begin{subfigure}[b]{0.8\textwidth}
        \centering
        \includegraphics[width=\textwidth]{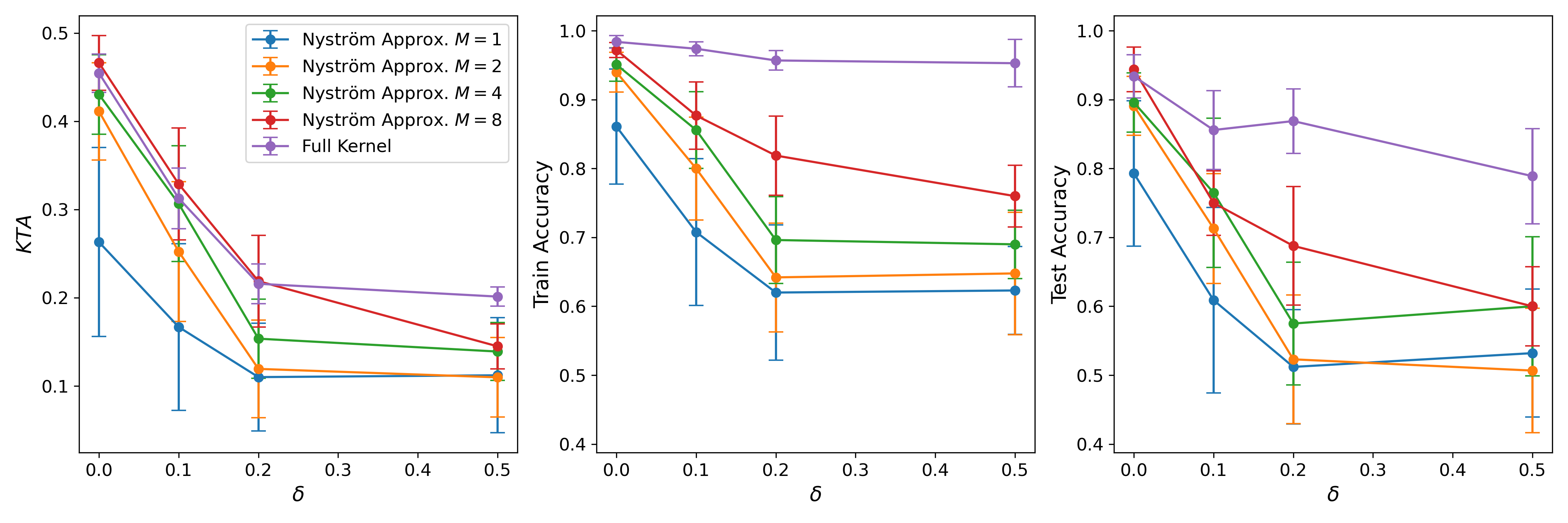}
        \caption{}
        \label{fig:cohr2}
    \end{subfigure}
    \caption{Performance of the Standard approach VS the Nyström approximation with different $Ms$ for an increasing strength of coherent noise $\sigma$ in the checkers (Fig.\ref{fig:cohr1}) and corners dataset (Fig.\ref{fig:cohr2}).}
    \label{fig:coherent_noise}
\end{figure*}

\section{\uppercase{Numerical Results}}

In this section,  we empirically compare the Nyström approximation method and the standard method on a set of simple binary 2D classification tasks. The datasets selected include the \emph{checkers} and \emph{donuts} datasets from \cite{hubregtsen2022training}, a manually created \emph{corners} dataset, and the \emph{spirals} dataset generated using the \emph{make\_moons} method from \emph{scikit-learn}. For a more detailed explanation, please refer to Appendix \ref{appendix}.

To compare the methods, we will use three metrics: the KTA between the current quantum kernel and the ideal kernel for the entire training dataset, the training accuracy, and the testing accuracy. Note that calculating the first two metrics requires computing the training kernel matrix, while the last metric requires computing the testing kernel matrix. This is where the standard approach and the Nyström approximation differ, as previously explained. Due to the inherent randomness of parameter initialization, results are averaged over $10$ seeds and plotted alongside the standard deviation.

For the experiments, the input scaling weights are initialized as $1$s, ensuring they initially have no effect on the output of the quantum kernel, and the variational weights sampled uniformly between $-\pi$ and $\pi$. The methods were implemented using PyTorch and PennyLane and all results obtained using state-vector simulators. The parameters were optimized using the ADAM optimizer with a learning rate of $0.1$ and the ansatz had a constant depth of $5$ layers. Moreover, during the training iterations, we always used a mini-batch size $D=8$.

\subsection{\uppercase{Method Comparison on Noiseless Simulator}}
\label{noiseless_numerical}

We will start by analyzing both methods in ideal conditions, that is, in the absence of noise.

The performance comparison between the standard approach and the Nyström approximation on four benchmark datasets is illustrated in Fig.\ref{fig:noiseless_results}.

As observed in the KTA graphs, the standard approach consistently achieves the highest KTA across all four datasets. Moreover, a lower value of $M$ typically results in a reduced KTA for the Nyström approximation method. This is expected, as lower values of $M$ lead to less accurate approximations of the training and testing kernel matrices. However, this lower KTA does not necessarily imply reduced accuracies, as evidenced by the training and testing results. In some datasets, the Nyström method achieves both training and testing accuracies of 100\%, despite the lower KTA. 

These findings suggest an important consideration: if one is willing to potentially sacrifice a slight amount of accuracy for a significantly reduced number of quantum circuit executions, the Nyström Approximation should be considered. Moreover, it's also clear that the training approach from \cite{sahin2024efficient} of using a mini-batch of size $D$ every training step works, since in all datasets both accuracy and KTA increased throughout training.

However, some nuances should be discussed. It's not clear precisely how both $M$ and $D$ are related to the datasets' complexity and/or size. We used several training datasets with sizes ranging from $30$ to $100$ training data points and observed that, for all of them, $D=8$ and $M\in\{2,4,8\}$ lead to considerably high accuracies, achieving near or perfect classification on most of the datasets. However, for larger and more complex datasets, it may be necessary to increase both the values of $D$ and $M$. It may also be helpful to select the $D$ data points and $M$ landmarks not randomly but instead based on more sophisticated techniques. For instance, the $M$ landmarks for the Nyström approximation technique can be selected using k-means clustering, ensuring that they are representative of the entire dataset, effectively better approximating the kernel matrix. However, testing such techniques is out of the scope of this work, which intends to simply demonstrate for the first time a quantum-kernel pipeline that depends only linearly on the training dataset size. In short, we do not claim that the values of $M$ and $D$ used in this work and the technique we used to select them (uniformly random sampling) are guaranteed to provide good results for any dataset. In fact, this pipeline should be highly dependent upon the dataset and, consequently, these hyperparameters should be carefully chosen.

These results are in ideal conditions, if one had access to a perfect quantum computer. In the next subsections we will look at how the methods compare under different types of noise.

\begin{figure*}[ht]
    \centering
    \begin{subfigure}[b]{0.8\textwidth}
        \centering
        \includegraphics[width=\textwidth]{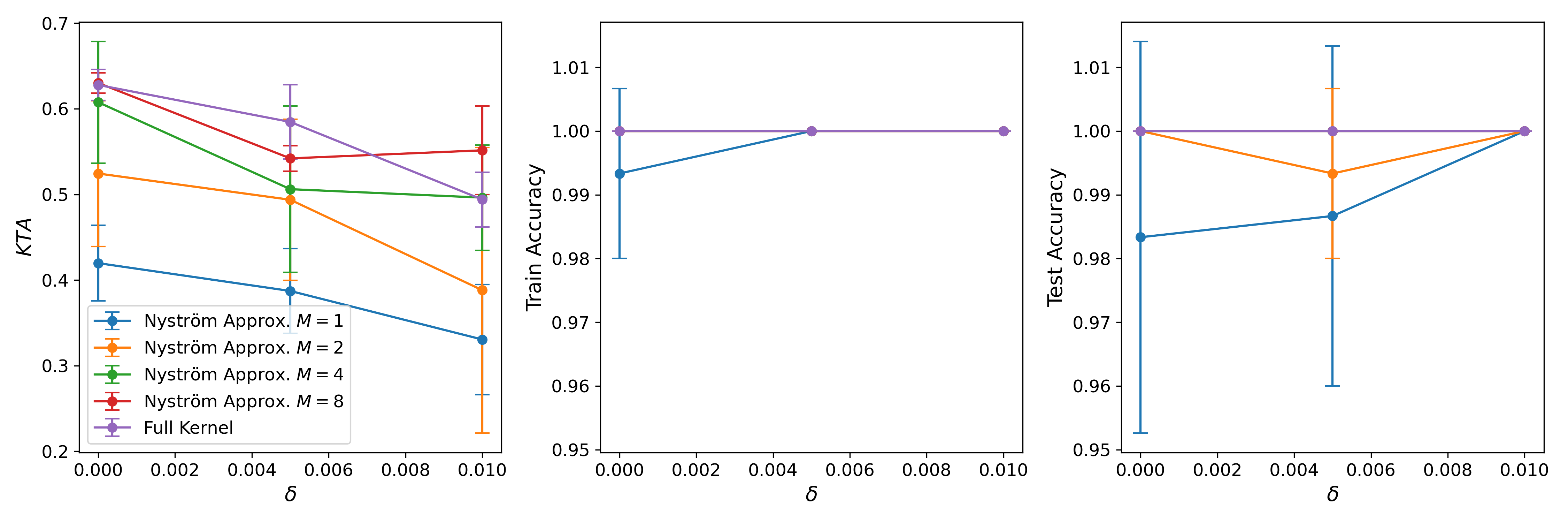}
        \caption{}
        \label{fig:depol1}
    \end{subfigure}
    \\[1em]
    \begin{subfigure}[b]{0.8\textwidth}
        \centering
        \includegraphics[width=\textwidth]{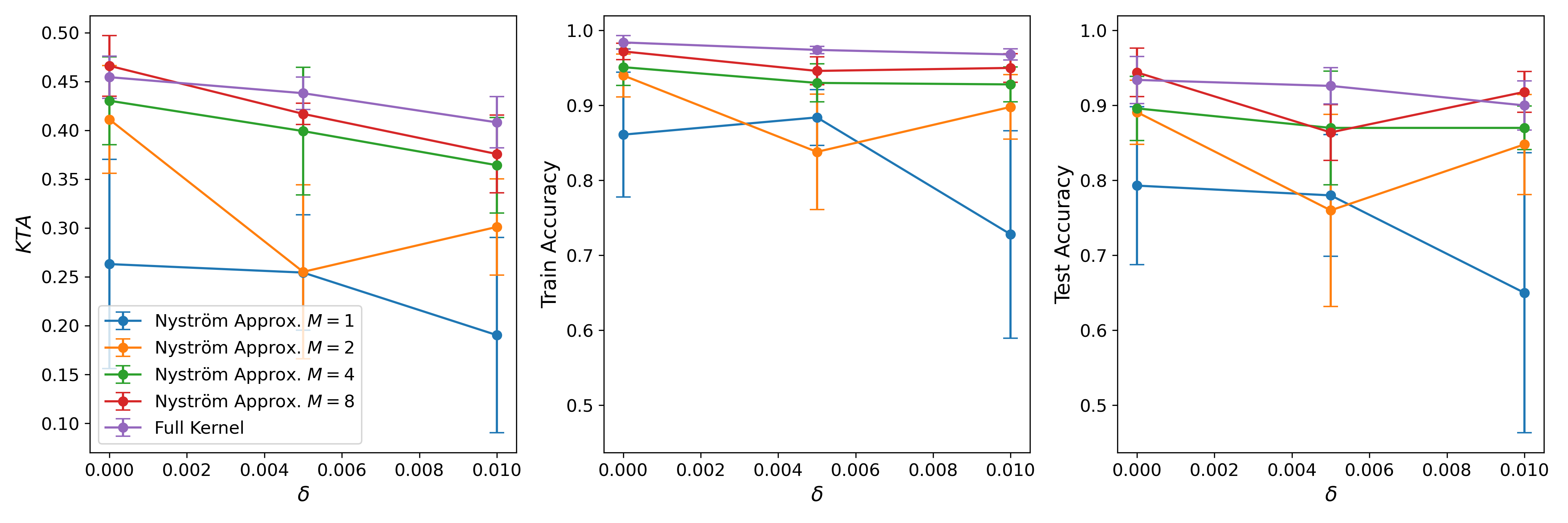}
        \caption{}
        \label{fig:depol2}
    \end{subfigure}
    \caption{Performance of the Standard approach VS the Nyström approximation with different $Ms$ for an increasing probability of depolarizing noise $p$ in the checkers (Fig.\ref{fig:depol1}) and corners dataset (Fig.\ref{fig:depol2}).}
    \label{fig:depolarizing_noise}
\end{figure*}

\subsection{\uppercase{Method comparison under Coherent Noise}}

Coherent noise are errors that preserve the unitary evolution of the quantum circuit but that still affect its output \cite{cai2020mitigating}. In \cite{skolik2023robustness}, coherent noise was modeled as under or over-rotations of the parametrized quantum gates, which can be seen as a miscalibration of the quantum gates. We will follow a similar approach. The variational quantum parameters $\theta$ will be changed according to $\theta\rightarrow\theta + \delta\theta$, where $\delta\theta$ are sampled according to a Gaussian distribution of mean $0$ and variance $\sigma^2$. The perturbations $\delta\theta$ are re-sampled if a new observable is being measured or if the circuit is being evaluated with a different set of parameters \cite{skolik2023robustness}.

The performance of both the standard approach and the Nyström approximation under increasing strength of coherent noise $\sigma$ can be seen in Fig. \ref{fig:coherent_noise}.

Firstly, it is evident that the standard approach demonstrates better resistance to coherent noise. For both datasets, although the KTA decreases significantly as $\sigma$ increases, the training and testing accuracy only slightly decline for the corners dataset. In contrast, the Nyström approximation method shows a sharp decrease in KTA, training, and testing accuracy when $\sigma>0.2$. Furthermore, as observed in the noiseless setting, lower values of $M$ tend to perform worse. These results are unsurprising, as the Nyström method approximates the kernel matrix using only a limited number of quantum circuit executions. If these executions are noisy, it is expected that the resulting approximation will be even less accurate. In summary, the approximation and the noise compound, whereas the standard approach can tolerate more noise.

\subsection{\uppercase{Method Comparison under Depolarizing Noise}}

In this section, we aim to analyze the performance of both methods under a different type of noise: \emph{Depolarizing Noise}. This noise affects a quantum state by replacing it with a mixed state with probability \(p\), or leaving it unchanged otherwise. It's important to highlight that, in the presence of device noise, the adjoint operation required to compute the quantum kernel function may not be feasible. However, as shown in \cite{hubregtsen2022training}, it is indeed possible in the case of depolarizing noise.

In Pennylane, depolarizing noise can be implemented using the \emph{Depolarizing Channel} operation. We model this type of noise by introducing single-qubit depolarizing channels after each quantum gate.

The performance of both methods under depolarizing noise as a function of the probability $p$ can be seen in Fig.\ref{fig:depolarizing_noise}.

Here too, similar results can be observed. The standard approach is more robust against depolarizing noise when compared with the Nyström method. Nonetheless, both methods are able to effectively reach high training and testing accuracies. In particular, if one uses the Nyström method with a larger $M$, then the results are at least comparable to the standard approach, while still requiring substantially less quantum circuit executions.

\section{\uppercase{Conclusion}}

In this work we adapted a low-rank matrix approximation method named Nyström approximation to quantum embedding kernels, effectively reducing the number of quantum circuit executions required to construct the training and testing kernel matrices. In doing so, we also introduce the first quantum kernel pipeline that has an end-to-end quantum complexity that depends only linearly on the training dataset size $N$. Other works had already introduced a linear complexity during the training stage, but not during the training matrix creation. Furthermore, we show that, in the noiseless scenario, an SVM trained from a quantum kernel using the Nyström approximation has a performance comparable to that of an SVM trained using the standard quantum kernel approach, without any approximation. Finally, we compare both methods in the presence of coherent and depolarizing noise and empirically demonstrate that both methods perform well under realistic levels of noise.

\section*{\uppercase{Code Availability}}

The source code for replicating the results is available on Github at \url{https://github.com/RodrigoCoelho7/qekta}.

\section*{\uppercase{Acknowledgements}}

The research is part of the Munich Quantum Valley, which is supported by the Bavarian state government with funds from the Hightech Agenda Bayern Plus.

\bibliographystyle{apalike}
{\small
\bibliography{example}}

\section*{\uppercase{Appendix}}\label{appendix}

\begin{figure}[h]
    \centering
    \begin{tabular}{cc}
        \begin{subfigure}[t]{0.48\columnwidth}
            \centering
            \includegraphics[width=\textwidth]{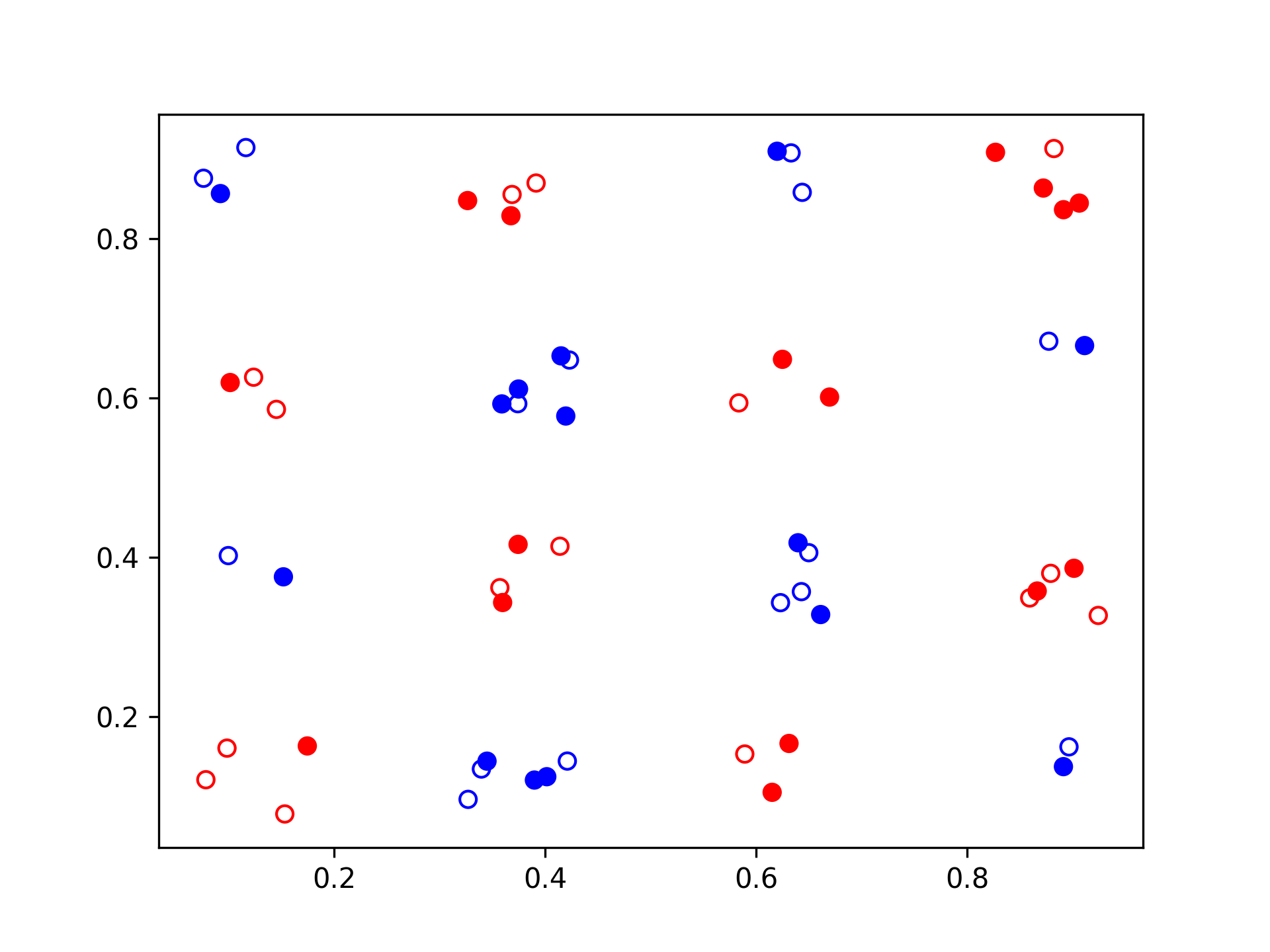}
            \caption{Checkers Dataset}
        \end{subfigure} &
        \begin{subfigure}[t]{0.48\columnwidth}
            \centering
            \includegraphics[width=\textwidth]{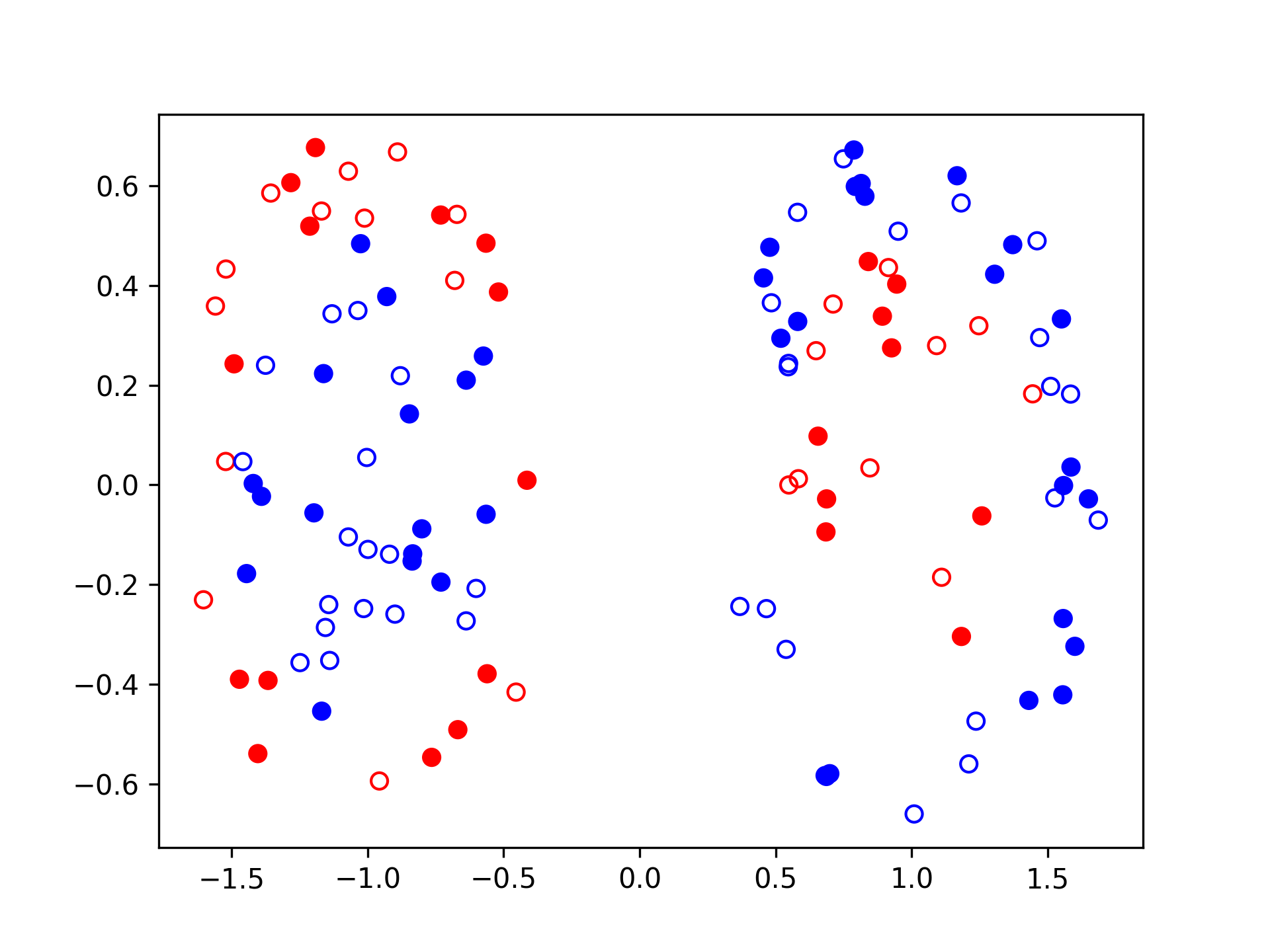}
            \caption{Donuts Dataset}
        \end{subfigure} \\
        \begin{subfigure}[t]{0.48\columnwidth}
            \centering
            \includegraphics[width=\textwidth]{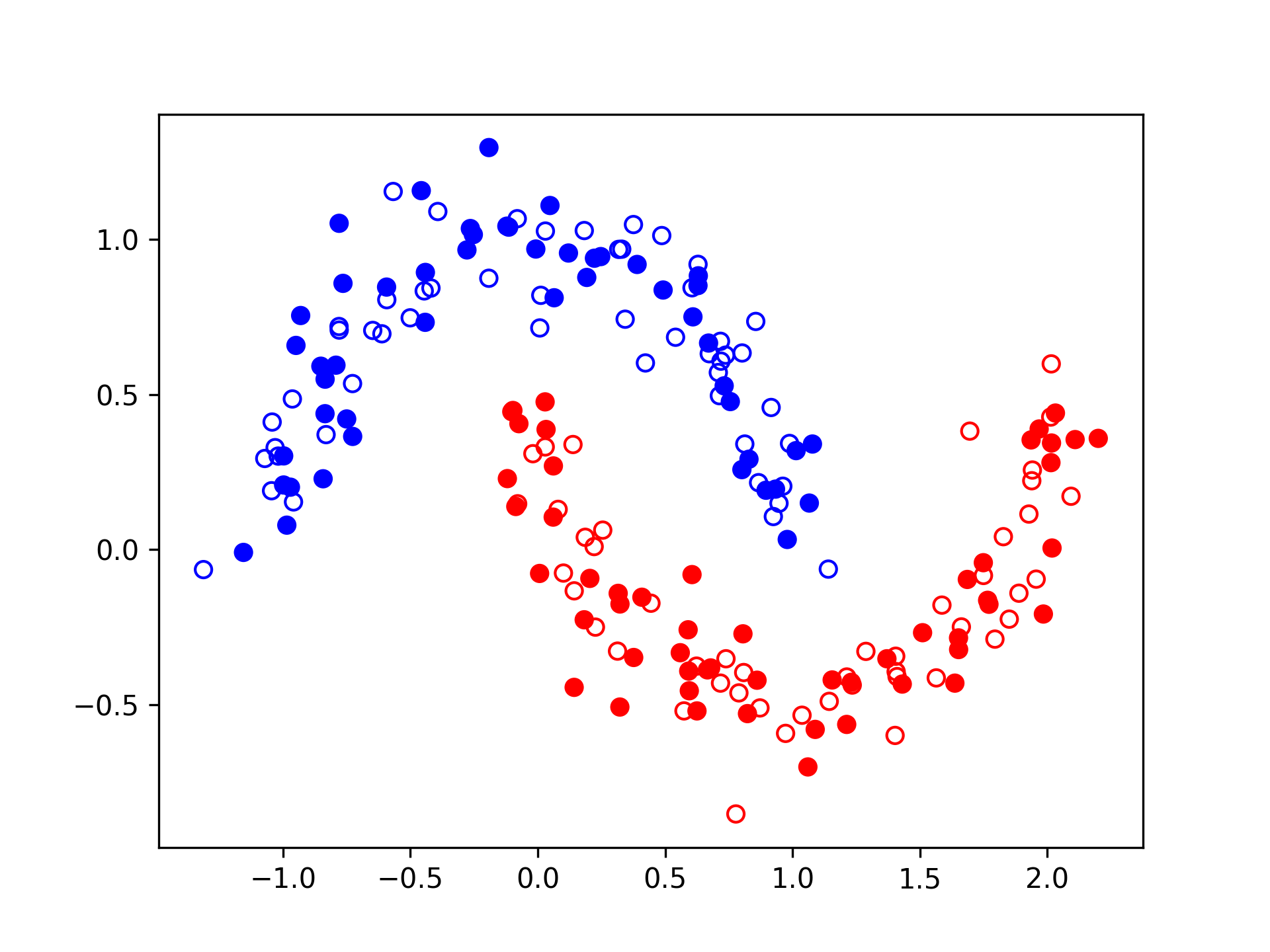}
            \caption{Spirals Dataset}
        \end{subfigure} &
        \begin{subfigure}[t]{0.48\columnwidth}
            \centering
            \includegraphics[width=\textwidth]{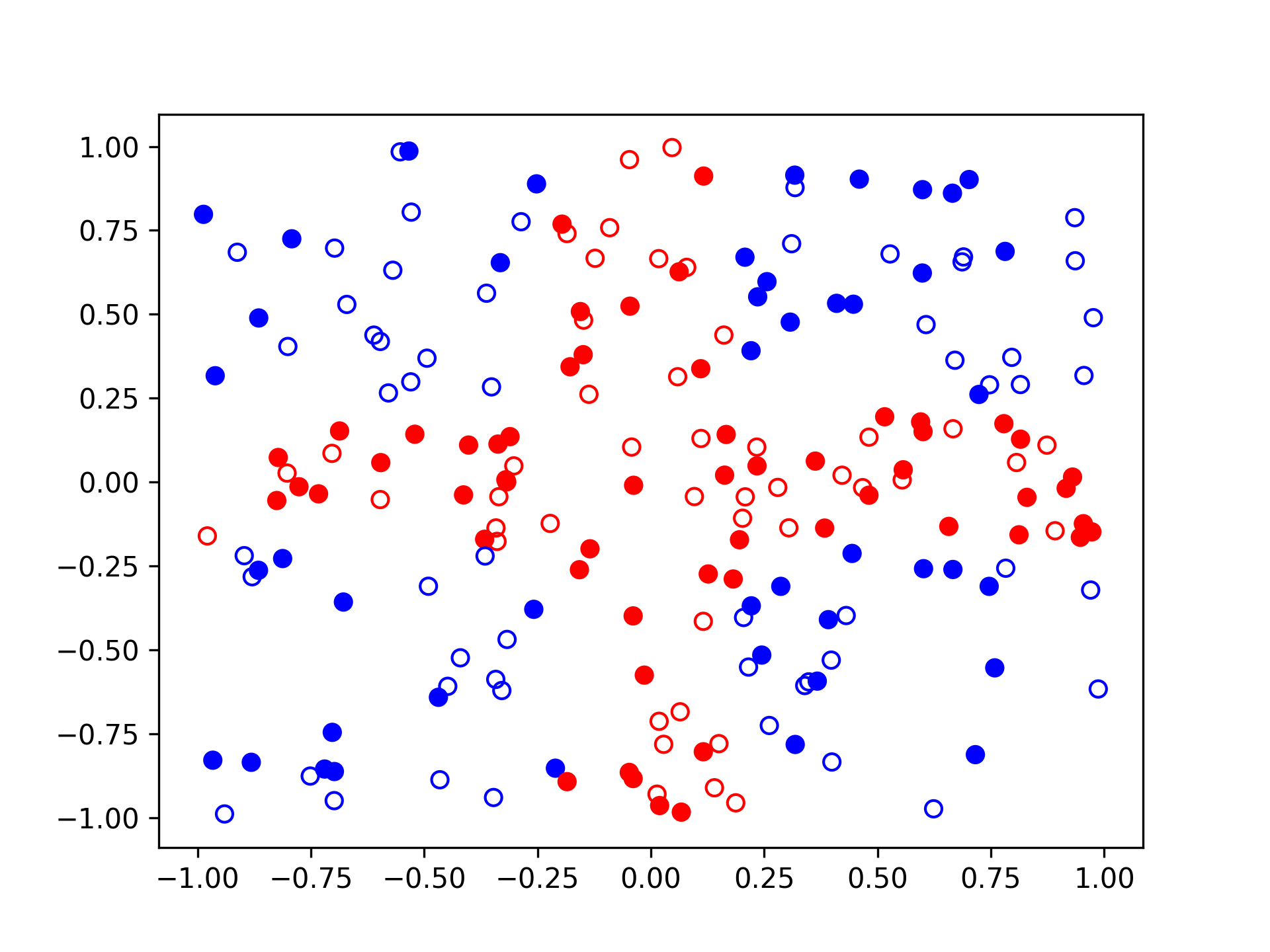}
            \caption{Corners Dataset}
        \end{subfigure} 
    \end{tabular}
    \caption{Red and blue represent the two classes. Circles/Circumferences represent training/testing data points.}
    \label{fig:datasets}
\end{figure}

The four binary 2D datasets used in this work for benchmarking the models can be seen on Fig.\ref{fig:datasets}. A specification of the datasets and how to create them:
\begin{itemize}
    \item \emph{Checker} from \cite{hubregtsen2022training}. This dataset consists of $30$ training and $30$ testing data points. Used on both noiseless and noisy numerical results. The reader is referred to \cite{hubregtsen2022training} for more details on how to create it.
    \item \emph{Donuts} from \cite{hubregtsen2022training}. This dataset consists of $60$ training and $60$ testing data points. Used on noiseless numerical results. The reader is referred to \cite{hubregtsen2022training} for more details on how to create it.
    \item \emph{Spirals} from \emph{scipy's make\_moons} method. This dataset consists of $100$ training and $100$ testing data points. Used on noiseless numerical results.
    \item \emph{Corners}. This dataset consists of $100$ training and $100$ testing data points. To generate the dataset, we first define a square of size $2$ (both x and y-axis are between $-1$ and $1$) and randomly sample a set of points within this region. Each point is then classified based on its location relative to four quarter circles, each centered at one of the corners of the square. Specifically, if a point falls within any of these quarter circles, it is labeled as $-1$; otherwise, it is labeled as $1$. Used on noiseless and noisy 
\end{itemize}

\end{document}